\documentclass[11pt,a4paper]{article}
\usepackage{jcappub}
    
\usepackage{dcolumn}
\usepackage{bm}
\usepackage{color}
\usepackage{epsf}
\usepackage{graphicx}
\usepackage{amsmath}
\usepackage[utf8]{inputenc}
\usepackage[dvipsnames]{xcolor}
\usepackage{dcolumn}
\usepackage{physics}

\hypersetup{colorlinks,linkcolor={blue},citecolor={teal},urlcolor={violet}}

\newcommand{\INFN}{INFN - Sezione di Napoli, Complesso Univ. Monte S. Angelo, I-80126 Napoli, Italy}
\newcommand{\UNI}{Dipartimento di Fisica "Ettore Pancini", Universit\'a degli studi di Napoli "Federico II", Complesso Univ. Monte S. Angelo, I-80126 Napoli, Italy}

\title{IceCube constraints on Violation of Equivalence Principle}

\author[a,b]{Damiano F.G. Fiorillo}
\author[a,b]{Gianpiero Mangano}
\author[a,b]{Stefano Morisi}
\author[a,b]{Ofelia Pisanti}

\affiliation[a]{\UNI}
\affiliation[b]{\INFN}

\emailAdd{dfgfiorillo@na.infn.it}
\emailAdd{mangano@na.infn.it}
\emailAdd{stefano.morisi@na.infn.it}
\emailAdd{pisanti@na.infn.it}

\date{\today}

\abstract{Among the information provided by high energy neutrinos, a promising possibility is to analyze the effects of a Violation of Equivalence Principle (VEP) on neutrino oscillations. We analyze the recently released IceCube data on atmospheric neutrino fluxes under the assumption of a VEP and obtain updated constraints on the parameter space with the benchmark choice that neutrinos with different masses couple with different strengths to the gravitational field. In this case we find that the VEP parameters times the local gravitational potential at Earth can be constrained at the level of $10^{-27}$. We show that the constraints from atmospheric neutrinos strongly depend on the assumption that the neutrino eigenstates interacting diagonally with the gravitational field coincide with the mass eigenstates, which is not \textit{a priori} justified: this is particularly clear in the case that the basis of diagonal gravitational interaction coincide with the flavor basis, which cannot be constrained by the observation of atmospheric neutrinos. Finally, we quantitatively study the effect of a VEP on the flavor composition of the astrophysical neutrinos, stressing again the interplay with the basis in which the VEP is diagonal: we find that for some choices of such basis the flavor ratio measured by IceCube can significantly change.}

\begin{document}

\maketitle
\flushbottom


\section{Introduction}

The recently rising field of neutrino astronomy has the potential to deepen our understanding of many open problems in fundamental physics. On one side, the observation of extraterrestrial neutrinos may provide key information on the determination of the astrophysical origin of cosmic rays; on the other hand, the high energies of atmospheric and astrophysical neutrinos renders them an ideal test case for the identification of effects of physics Beyond Standard Model (BSM). Extensions of the Standard Model may in fact be responsible for changes in their energy spectrum, arrival times and directions, and flavor structure. Among the latter, a significant example is represented by the existence of Violations of the Equivalence Principle (VEP) in gravitational interactions.

The Equivalence Principle lies at the very foundation of the theory of gravitation, that is, general relativity. In its weak form, it states that all test points follow identical trajectories in the same gravitational field: this can be cast in the assumption that  the effects of a gravitational field are locally equivalent to the effects seen in an accelerated observer frame. The Weak Equivalence Principle requires, therefore, the equality of inertial and gravitational masses, which implies that all particles couple to the gravitational field with the same coupling strength, parametrized by the Newton constant $G_N$. The observation of a VEP would constitute a clear indication towards a change in our understanding of gravity: since we are aware that the theory of gravity as it stands is not complete in the ultraviolet regime, as it is still a classical theory which cannot be promoted to a quantum renormalizable model, the investigation of VEP may therefore be a crucial test of fundamental importance towards the construction of a new model. For this reason, the Weak Equivalence Principle is still nowadays tested by a variety of experiments including torsion-balance experiments \cite{Wagner:2012ui}, motion of bodies in the solar system \cite{Overduin:2013soa}, spectroscopy of atomic levels \cite{Hohensee:2013cya} and pulsars \cite{Damour:1991rq,Horvat:1998st,Barkovich:2001rp}. 

Another possibility of probing VEP, as mentioned above, is the study of high energy neutrinos. In the standard paradigm of neutrino oscillations, the eigenstates of free propagation do not coincide with the flavor eigenstates: the difference in neutrinos rest mass cause a relative dephasing between the propagation eigenstates which leads to the observed oscillations. In the presence of VEP a similar effect might be expected. In fact, if different combinations of flavor eigenstates couples with different intensities to an external gravitational potential, the induced relative dephasing would produce a neutrino oscillation behaviour quite different from the conventional ones. Actually, this mechanism was originally proposed in order to explain the solar neutrino problem \cite{Pantaleone:1992ha,Butler:1993wi,Bahcall:1994zw,Halprin:1995vg,Mureika:1995ap,Mureika:1996de,Mureika:1996ud,Mansour:1998nb,Gago:1999hi,Casini:1999kt,Majumdar:2000sd}, before the establishment of the conventional oscillation framework. The agreement of the standard scenario of neutrino oscillations with all the experimental data suggests a change of perspective: the observation of neutrino oscillations can now be used to constrain the existence of VEP. Along these lines, observations conducted on neutrinos from many sources have led to competitive constraints  \cite{Gasperini:1988zf,Gasperini:1989rt,Minakata:1994kt,Valdiviesso:2011zz,Foot:1997kk,Foot:1998pv,Fogli:1999fs,GonzalezGarcia:2004it,GonzalezGarcia:2005xw,Battistoni:2005gy,Morgan:2007dj,Abbasi:2009nfa,Pakvasa:1988gd,Guzzo:2001vn,Minakata:1996nd,Iida:1992vh,Mann:1995nw,Esmaili:2014ota,Diaz:2020aax}.

From a simple dimensional analysis, the inverse wavelength of the VEP induced oscillations can only be proportional to the energy of the neutrino. An ideal source of data able to constrain VEP is therefore provided by the neutrino telescopes \cite{Aslanides:1999vq,Aartsen:2014oha,Barwick:2007vba,Meures:2012fka,Martineau-Huynh:2015hae, Schulz:2009zz,Blaufuss:2015muc,Collaboration:2011nsa,Belolaptikov:1997ry,Riccobene:2017fpr,Kappes:2007ci,Goldschmidt:2001qd,Ahrens:2002dv}. In this work we focus on the atmospheric neutrinos detected by IceCube in the energy range between $100$ GeV and $1$ PeV roughly. In this energy range, conventional oscillations should produce no effects on the fluxes, since the wavelength of oscillation is larger than the Earth radius. On the other hand, VEP induced oscillations can significantly change this conclusion. We study the effects of VEP on atmospheric fluxes, and we use the data released in 2018 from IceCube to provide updated constraints on the VEP parameter space. In doing so, we update the results presented in Ref. \cite{Esmaili:2014ota}, which were already stronger than the ones obtained by studies of atmospheric neutrinos with other experiments \cite{GonzalezGarcia:2004wg,Battistoni:2005gy,Abbasi:2009nfa}; with the new data released by IceCube, we are able to lower the constraints on the VEP strengths by even an order of magnitude.

As a benchmark choice we assume, similarly to what is done in the literature, that neutrinos belonging to different mass eigenstates couple with different strengths to the gravitational field. There is however no \textit{a priori} reason why the mass eigenstates should coincide with the basis in which the gravitational field is diagonal, which we will name throughout this work gravitational basis.  Even though a full study of the constraints on the strength of the VEP for all possible choices is beyond the purpose of this work, we discuss which kind of qualitative changes, arguing that for some specific choices the effects on the atmospheric neutrinos might even vanish: in these cases, we expect that no constraints on the model can be derived by analysis of the atmospheric flux.


As a further interesting issue we analyze the effects that VEP might induce on the flavor composition of cosmic neutrinos. Similarly to Ref. \cite{Minakata:1996nd}, we propose that the relative dephasing acquired by neutrinos in the gravitational fields felt on their way to Earth might disrupt the conventional oscillation scenario. However, differently from Ref. \cite{Minakata:1996nd}, we focus on a scenario which is substantially independent of the specific spatial structure of the gravitational fields, and which can therefore be analyzed in full generality, if VEP is sufficiently strong. In this scenario, we find that for different choices of the gravitational basis the flavor composition of cosmic neutrinos can be drastically changed by the presence of VEP: we compare our predictions with the current IceCube limits on the flavor composition, finding that for some choices of the gravitational basis a tension is already present with the data. We also analyze the relation with the sensitivity of the future experiment IceCube-Gen2, emphasizing the complementarity between the information that atmospheric and astrophysical neutrinos can provide. 


\section{Neutrino oscillations in the presence of VEP} \label{sec:oscillations}

Violations of Equivalence Principle have long been known to influence the oscillation properties of neutrinos during their propagation in gravitational potentials; in this Section we describe the general formalism used to obtain the oscillation properties in such a condition. Throughout this work we use units in which $\hbar=c=1$.

Neutrino oscillations can be generally understood as the dephasing of neutrino propagation eigenstates caused by their different masses. If neutrinos propagate through a dense environment of electrons with number density $N_e (\mathbf{r})$ a further source of dephasing is due to the coherent  forward scattering on electrons. The equations describing the propagation along a path length $l$ of a neutrino state $\ket{\psi}=\sum_{\alpha} c_\alpha \ket{\nu_\alpha}$, with $\alpha=e,\; \mu,\; \tau$, are
\begin{equation}\label{eq:propanovep}
    i\frac{dc_\alpha}{dl}=\sum_{j,\beta} U_{\beta j} U^*_{\alpha j} \frac{\delta m_j^2}{2E} c_\beta + V(\mathbf{r}) \delta_{\alpha e} c_e,
\end{equation}
where $E$ is the neutrino energy, $U_{\alpha j}$ are the components of the PMNS matrix, $\delta m_j$ are the neutrino mass splittings, assuming $\delta m_1^2=0$. Furthermore, $V(\mathbf{r})=\sqrt{2} G_F N_e(\mathbf{r})$ is the matter potential. 

In this context a VEP could make the coupling between neutrinos and an external gravitational potential different for different neutrino states. In other words, among the vector space spanned by the neutrino flavor states, three linear combinations will be selected which couple diagonally with the gravitational field. Such a basis $\ket{\tilde{\nu}_a}$ is defined by a new unitary matrix $\tilde{U}$ as
\begin{equation}
    \ket{\nu_\alpha}=\sum_a \tilde{U}_{\alpha a} \ket{\tilde{\nu}_a}.
\end{equation}
In the presence of a weak gravitational potential $\phi$ the spacetime metric felt by each of these neutrino eigenstates will be different: to first order in $\phi$ we can write it as $g_{\mu \nu}=\eta_{\mu \nu}+2\gamma_a \phi\delta_{\mu\nu}$, where $\eta_{\mu\nu}$ is the Minkowski metric (we use the ($+ - - -$) signature), $\delta_{\mu\nu}$ is the Kronecker delta and $\gamma_a$ are three coefficients measuring the intensity of the coupling of the gravitational field with each state. In the absence of VEP all these parameters are equals, $\gamma_a=1$. In the presence of VEP one of the $\gamma_a$ can be still set equal to unity, by a redefinition of the gravitational potential.

The propagation equations (\ref{eq:propanovep}) are then changed to \cite{Valdiviesso:2011zz}
\begin{eqnarray}\label{eq:propavep}
i\frac{dc_\alpha}{dl}=\sum_{j,\beta} U_{\beta j} U^*_{\alpha j} \frac{\delta m_j^2}{2E} c_\beta + V(\mathbf{r}) \delta_{\alpha e} c_e
 +2E\phi\sum_{a,\beta} \tilde{U}_{\beta a} \tilde{U}^*_{\alpha a} \gamma_a c_\beta.
\end{eqnarray}
Just as in the case of conventional oscillations, neutrino propagation will only be sensitive to the difference between the phases induced by VEP: this can be explicitly proven by subtracting a term $2E\gamma_1 c_\alpha$ from Eq. (\ref{eq:propavep}), after which the equation becomes only dependent on the quantities $\gamma_{21}=\gamma_2-\gamma_1$ and $\gamma_{31}=\gamma_3-\gamma_1$. In the following we will express our results in terms of these two quantities only; we will also generally refer to them as $\gamma_{ij}$.

The choice of the matrix $\tilde{U}$ is in principle completely arbitrary, and amounts to selecting the basis of diagonal coupling with the gravitational field: as we said already, throughout this paper we will call this gravitational basis. Different choices in this regard can cause drastically different effects on neutrino oscillations: for example, if $\tilde{U}=U$, that is the gravitational basis coincides with the mass basis, VEP will have the same effect as a modified dispersion relation. We focus on this latter choice, as in Ref. \cite{Esmaili:2014ota}. We will comment in the following on the observable effects for a different basis.

We focus on the effects of a VEP on atmospheric neutrinos propagating through the Earth. We adopt the Preliminary Reference Earth Model (PREM) \cite{Dziewonski:1981xy} for describing the electron number density. The dominant source of the gravitational potential at the Earth is believed to be the Great Attractor, resulting in $\phi\sim -3\times 10^{-5}$ \cite{Kenyon:1990ve}, though uncertainties on this value remain. In order to provide definite results, we will use throughout this work the combination $\gamma_a \phi$, as in Refs. \cite{Esmaili:2014ota,Guzzo:2001vn,Valdiviesso:2011zz,GonzalezGarcia:2004wg,Battistoni:2005gy}. Finally, we use the best fit values for the oscillation parameters obtained in Ref. \cite{Zyla:2020zbs} (see also \cite{Fogli:2013faa,deSalas:2017kay,Esteban:2016qun}). It is noteworthy that the spatial dependence of the gravitational potential generated by the Earth can be completely neglected, since the latter is of the order of $\phi_{\text{Earth}}\sim 2G_N M_{\text{Earth}} R^{-1}_{\text{Earth}}\sim 10^{-9}$, and is therefore negligible compared with the background potential generated by the Great Attractor.

By numerically solving the coupled system of Eqs. (\ref{eq:propavep}) we can determine the probability that a neutrino entering with flavor $\alpha$ escapes the Earth in the flavor eigenstate $\beta$. This probability, which we denote by $P_{\alpha\beta}$, will be a function of the neutrino energy and declination, as well as of the VEP parameters $\gamma_{21}$ and $\gamma_{31}$.

While the numerical solution of Eqs. (\ref{eq:propavep}) is able to provide the result in a reasonable computation time, it is still useful to get an approximate analytical understanding of the behavior of the solution. The three terms in the propagation equation define three types of characteristic lengths: the conventional oscillation lengths $E\delta m_i^{-2}$, a matter length $\overline{V}^{-1}$, where $\overline{V}$ is an average of $V(\mathbf{r})$ inside the Earth, and the VEP oscillation lengths $(E\gamma_{ij}\phi)^{-1}$. The dominant effect will be produced by the process with the shortest characteristic length. Due to their behavior with energy, we expect the conventional oscillation effects to dominate at low energies. 

At an energy of about $100$ GeV, the conventional oscillation lengths are both larger than the Earth diameter, so that no oscillations are expected: on the other hand, since the VEP oscillation length decreases with energy, it is just in this energy range that its effect can become substantial and perhaps observed or strongly constrained. At energies so high that the VEP processes dominate over the matter term, extremely fast oscillations will be set up: the function $P_{\alpha\beta}$ will therefore be extremely rapidly changing. Its average value can be estimated, in the same way as it is done for conventional oscillations , as
\begin{equation}\label{eq:average}
    P_{\alpha\beta}=\sum_a |\tilde{U}_{\alpha a}|^2 |\tilde{U}_{\beta a}|^2.
\end{equation}
At intermediate energies the interplay between matter effects and VEP dephasing can give rise to slow coherent oscillations which do not average out. In this regime, analytical methods can only determine the solution for simplified models of matter densities.

\section{Atmospheric neutrinos at IceCube} \label{sec:atmoic}

Since the effect of VEP grows with the energy, an ideal laboratory for testing it is to look at high energy neutrinos. In particular, for $\gamma_{ij} \phi\,\sim 10^{-26}$, VEP induced oscillations are increasingly felt for energies above 100 GeV. In this energy range the dominant source of neutrinos detected at Earth is the collision of cosmic ray particles in the atmosphere. The resulting atmospheric neutrinos propagate from the top of the atmosphere to the Earth surface: many experiments have been able to detect and characterize these neutrino fluxes. For this work, we focus on the IceCube experiment: even though its primary aim is the detection of astrophysical neutrinos yet, IceCube collected during the last ten years a large number of data on atmospheric neutrinos in the range between 1 GeV and 100 TeV. These data can impose considerable constraints on a possible VEP, as we will show below.

In the following analysis we will mainly focus on upgoing atmospheric neutrinos which come from the Northern hemisphere and are detected by IceCube after crossing the Earth. In fact, downgoing atmospheric neutrinos have a shorter path, shorter than the typical VEP oscillation lengths.  Furthermore, these neutrino samples are strongly contaminated by the background of atmospheric muons. On the other hand, upgoing events from the Northern hemisphere are nearly completely composed of atmospheric muon neutrinos which pass through the Earth. Their baselines are therefore typically of the order of $7000$ km, which for $\gamma_{ij} \, \phi\sim 10^{-27}$ can be comparable or even much larger than the VEP induced oscillation lengths. 

In Ref. \cite{Esmaili:2014ota} an analysis of the IceCube data published at the time on atmospheric neutrinos was performed to obtain 90\% confidence level exclusion regions for the parameter space of this model of VEP. The data set analyzed in this work were the IC40 and the IC79 data samples: in particular, for the latter the effective area had not been provided by IceCube, so that a scaling argument was used to obtain it from the effective area for IC40.

During the last years IceCube has provided new measurements of the atmospheric fluxes, both in the IC79 configuration and in the final IC86 configuration. We therefore analyzed the newly released IceCube data in Ref. \cite{Aartsen:2016oji} under the assumption of a VEP; in particular, we focus on the IC79 data sample after one year of observation and on the IC86 data sample in 2011, henceforth denoted as IC86-11. Recently \cite{Aartsen:2019fau}, IceCube also provided an updated analysis of the track events detected in the energy range roughly above 10 GeV. These data have been fully released in Ref. \cite{Abbasi:2021bvk}, accounting for a lifetime of the experiment of roughly six years from 2012 to 2018, and are composed of 760923 events. The corresponding data sample will henceforth be denoted as IC86-12/18.

All three data samples provide the track events identified by IceCube: the latter are produced by the charged current interaction of muon neutrinos and antineutrinos, which generates muons and antimuons emitting Cherenkov radiation during their propagation. For this reason, tracks allow a rather precise angular reconstruction. On the other hand, only a moderate precision can be achieved in the energy reconstruction. In this work we focus only on an angular analysis, so that this does not constitute a severe drawback to our study.

The three data samples of IC79, IC86-11 and IC86-12/18 are composed respectively of 48362, 61313 and 760923 upgoing neutrino events with a zenith angle between $90^\circ$ and $180^\circ$. In all cases we binned the data in ten bins in $\cos\theta$, where $\theta$ is the zenith angle, from 0 to -1. The expected number of events per bin in the observation time $T$ as a function of $\gamma_{21}$ and $\gamma_{31}$ can be written as
\begin{eqnarray}
    N^{\text{th}}_i(\gamma_{21},\gamma_{31}) = \int dE \int_{\cos\theta_i}^{\cos\theta_{i+1}} d\cos\theta\, 2\pi \nonumber \\ \sum_{\alpha} \Phi_{\alpha}(\cos\theta, E) P_{\alpha\mu}(\cos\theta, E,\gamma_{21},\gamma_{31}) A_{\text{eff}} (\cos\theta,E) T,
\end{eqnarray}
where $A_{\text{eff}}$ is the effective area for the experiment of interest, both provided for IC79, IC86-11 and IC86-12/18 by IceCube together with the corresponding dataset, $\Phi_\alpha$ is the atmospheric flux, see \cite{Honda:2006qj}. $T$ is respectively of 316, 333 and 2198.2 days for IC79, IC86-11 and IC86-12/18. 

The integral over the energy is made over the whole energy range of the experiment and for this reason, our analysis is only weakly sensitive to the spectral shape of the atmospheric fluxes. In particular, we have verified that the introduction of a prompt charm component modeled as in Ref. \cite{Enberg:2008te} does not significantly change the results, since it amounts to only a small addition of events compared to the much larger number of neutrinos expected from conventional atmospheric fluxes. For the same reason we do not expect our results to be significantly changed by the presence of a small contamination of astrophysical events at high energies.

Above $10$ TeV, a significant factor to take into account is the attenuation of neutrinos due to the incoherent collisions with nuclei in their propagation through Earth. While this phenomenon is in principle already taken into account in the effective areas provided by IceCube, the interplay between the incoherent attenuation and the coherent oscillations is non trivial  and could only be fully analyzed by solving the transport equations for the master evolution equation of the neutrino density matrix. However, since the neutrino cross sections are roughly independent of flavor in the energy range below $1$ PeV, we can deduce that attenuation does not enter the flavor evolution of neutrinos and therefore, we do not discuss it further.

We performed a statistical maximum likelihood analysis on each data sample \textit{via} minimization of the $\chi^2$  function \cite{Esmaili:2014ota}
\begin{eqnarray}
    \chi^2 (\gamma_{21}\phi, \gamma_{31} \phi,\alpha,\beta)=  \sum_i \frac{[N_i^{\text{data}}-\alpha(1+\beta (0.5+\cos\theta))N_i^{\text{data}}]^2}{\sigma_{i,\text{stat}}^2+\sigma_{i,\text{sys}}^2} + \frac{(1-\alpha)^2}{\sigma_\alpha^2}+\frac{\beta^2}{\sigma_\beta^2},
\end{eqnarray}
where $\alpha$ and $\beta$ take into account the systematic uncertainties on the normalization and the angular distribution respectively of the atmospheric flux and $N_i^{\text{data}}$ are the number of events per bin; the average values of $\alpha$ and $\beta$ are respectively 1 and 0 and the uncertainties on these values $\sigma_\alpha=0.24$ and $\sigma_\beta=0.04$ \cite{Honda:2006qj}. We consider a statistical uncertainty given by the Poisson estimate $\sigma_{i,\text{stat}}=\sqrt{N_i^{\text{data}}}$ and a systematic uncertainty $\sigma_{i,\text{sys}}=f N_i^{\text{th}}$. Since an independent estimate of the systematic uncertainty is not provided by the IceCube collaboration, we use a value for $f$ which guarantees agreement between the data and the atmospheric fluxes within $90\%$ confidence level, similarly to the approach of Ref. \cite{Esmaili:2014ota}. We treat $\alpha$ and $\beta$ as nuisance parameters and we marginalize over them, obtaining an effectively two-dimensional likelihood.

In principle, both $\gamma_{21}$ and $\gamma_{31}$ can take positive and negative values. However, in pure VEP oscillations, similarly to the case of pure neutrino oscillations, only the relative sign influences the mixing probabilities. In matter oscillations the absolute sign can in principle cause measurable differences: however, for the values of $\gamma_{ij}$ of interest to us matter effects are only important for low energies. For this reason, we find, in agreement with the results of Ref. \cite{Esmaili:2014ota}, that the exclusion contours depend only on the relative sign. We will therefore distinguish between a case of VEP direct ordering ($\gamma_{21}$ and $\gamma_{31}$ with the same sign) and VEP inverse ordering ($\gamma_{21}$ and $\gamma_{31}$ with opposite signs).
\begin{figure}
    \centering
    \includegraphics[width=0.45\textwidth]{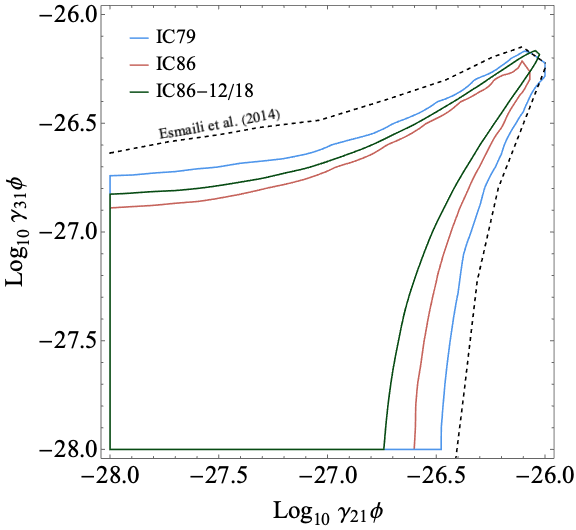}
    \includegraphics[width=0.45\textwidth]{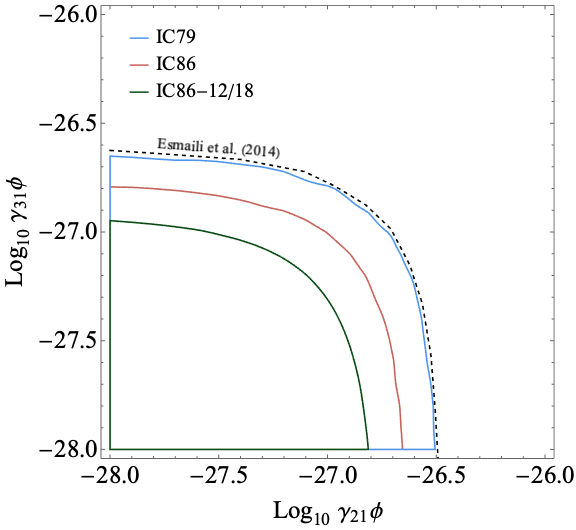}
    \caption{Allowed regions at 90\% confidence level in the $\gamma_{21}-\gamma_{31}$ plane for the IC79, IC86-11 and IC86-12/18 dataset shown in blue, orange and green respectively for the VEP direct (left) and inverse (right) ordering scenario. The black dashed curve is the 90\% confidence level exclusion contour obtained in Ref. \cite{Esmaili:2014ota}.}
    \label{fig:contatm}
\end{figure}
For the case of VEP direct ordering, we represent in the left panel of Fig. \ref{fig:contatm} the exclusion contours at 90\% confidence level in the plane $\gamma_{21}\phi-\gamma_{31}\phi$ for the three data samples used in our analysis. We also compare our exclusion contours with those obtained in Ref. \cite{Esmaili:2014ota} using the IC79 data and effective areas available at the time. For the case of VEP inverse ordering, we similarly represent in the right panel of Fig. \ref{fig:contatm} the exclusion contours at 90\% confidence level in the plane $\gamma_{21}\phi-\gamma_{31}\phi$, comparing with the exclusion region obtained in Ref. \cite{Esmaili:2014ota}.  

The results depend on the value of the systematic uncertainty, and a more conservative choice with a larger value may change the constraints: in this sense an independent estimate of such uncertainty might lead to more robust constraints. Nevertheless this study shows that for all data sets the constraints on VEP can be improved by even a factor of 2 compared to previous results in the literature. Surprisingly, the constraints from the 6-years data set IC86-12/18 are comparable with the ones from the 1-year data set IC86: this is due to the dominance of the systematic uncertainty over the statistical Poisson fluctuations. The systematic uncertainty grows proportionally to the number of events, so that there is no significant gain in increasing the number of events.


\section{Effects of a different gravitational interaction basis} \label{sec:basis}
The analysis of the atmospheric neutrinos at IceCube was performed under the assumption that the VEP is diagonal in the  mass eigenstates, meaning that neutrinos with different masses are subject to the gravitational interaction with different strengths. This assumption is quite reasonable, as gravity is expected to couple to the stress-energy tensor of matter, and as we said is mathematically equivalent to setting $\tilde{U}=U$. This can however, only be regarded as a benchmark choice, since there is no \textit{a priori} reason why the mass eigenstates, selected by the Yukawa coupling with the Higgs field, should coincide with the gravitational eigenstates selected by the VEP. A general analysis spanning the whole parameter space would of course be overwhelming in terms of computational time. In fact, the unitary matrix $\tilde{U}$ is parametrized in terms of three angles and one phase, if neutrinos are of Dirac nature.

Nonetheless, already at a qualitative level, some key features can be determined: in particular, if the gravitational basis coincide with the flavor basis, we expect no oscillation if the VEP effects dominate. In fact, in this extreme case propagation in a VEP dominated environment is diagonal in the flavor basis and only very weak constraints could be drawn from the atmospheric data sample. Actually, mild constraints could be determined in this case by an analysis of the lower energy data, as in Ref. \cite{Diaz:2020aax}, since in this energy region the interplay between conventional oscillations, which are diagonal in the mass basis, and VEP could produce observable effects. However, the typical values of $\gamma_{ij} \phi$ tested by these experiments are of the order of $10^{-24}$ \cite{Diaz:2020aax}, due to the fact that at lower energy the effects of VEP are weaker.

Since the effects of VEP grow with the neutrino energy, we consider the possibility that astrophysical neutrinos, the highest energy neutrinos known at present, could provide important information on this subject. During their propagation from the astrophysical sites of acceleration to the Earth, neutrinos travel through varying gravitational fields, leading to severe uncertainties on the potential $\phi$ to which they are subject. Coupled with the well known uncertainties on the nature of the astrophysical processes in which neutrinos are generated, this might seem to imply that there is no chance of disentangling VEP effects in astrophysical fluxes from the lack of information on their production and propagation. 

There is, however, a simple scenario in which definite conclusions can be drawn on the effects of VEP on the flavor composition of the astrophysical neutrinos. Since the conventional oscillation inverse wavelength is of the order of $\delta m_i^2 E^{-1}$, while the VEP oscillation inverse wavelength is of the order of $\gamma_{ij}\phi E$, we find that at an energy of 100 TeV already for values of $\gamma_{ij}\phi\sim 10^{-31}$ the VEP effects are dominant. Furthermore, the wavelengths connected with these oscillations can be at most of the order of $10^{10}$ m, which are much smaller than the cosmological distances of propagation. Therefore, regardless of the specific nature and spatial dependence of the gravitational potential felt by neutrinos on their way to the Earth, if $\gamma_{ij}\phi > 10^{-31}$ they will undergo extremely fast purely VEP oscillations. 

The requirement that the values of $\gamma_{ij}\phi$ should be so large throughout the path of the neutrinos does not seem to be an extreme one: as discussed in Ref. \cite{Minakata:1996nd}, in fact, the gravitational potential in the Intergalactic space, mostly caused by the Great Attractor, has typical values of $\phi\sim 5\times 10^{-6}$, comparable to those at the Earth. The values of $\gamma_{ij}$ which can be tested in this case are therefore of the same order of the ones testable \textit{via} analysis of atmospheric events: this scenario has the added advantage of being as model independent as possible, since the oscillations are so fast that they average out roughly independently of the specific spatial structure of the gravitational potential.

Under these conditions the mixing probabilities average out as in Eq. (\ref{eq:average}). The average values predicted depend on the structure of $\tilde{U}$: on the other hand, in the standard astrophysical scenario with no VEP, such probabilities depend on the PMNS matrix $U$. As we discussed above, the two matrices can be considerably different if the gravitational basis does not coincide with the mass basis. 

\begin{figure}
    \centering
    \includegraphics[width=0.7\textwidth]{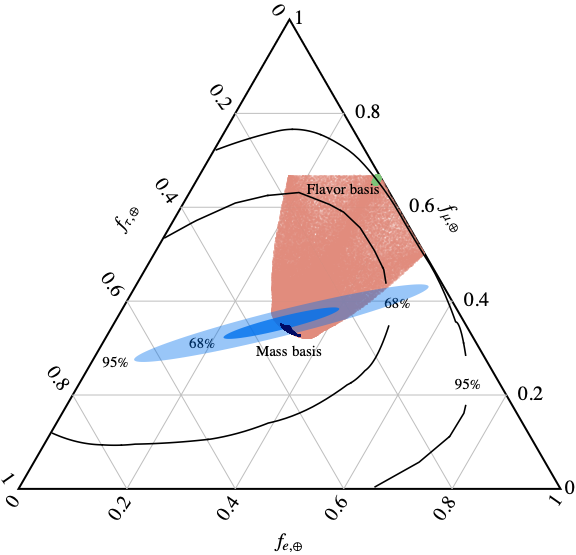}
    \caption{Flavor ratio at the Earth from a pion beam source after averaged VEP oscillations with $\gamma\phi>10^{-31}$: the orange region spans the whole possibilities for the gravitational basis; the green region corresponds to the case of gravitational basis coinciding with flavor eigenstates, while the black region corresponds to the case of gravitational basis coinciding with mass eigenstates. In the latter case we vary the oscillation parameters in the $3$ $\sigma$ intervals. The 68\% and 95\% confidence level obtained by IceCube \cite{Abbasi:2020zmr} are shown (black contours), as well as the projected flavor sensitivity of IceCube-Gen2 \cite{Bustamante:2019sdb} at 68\% and 95\% confidence level (blue ellipses).}
    \label{fig:flavtriang}
\end{figure}

While a detailed analysis of the IceCube data is beyond the aims of the present work, a simple example to test the relevance of this effect on the astrophysical data can be obtained by assuming that the astrophysical sources are in the pion beam regime, with a flavor composition at the source $(1:2:0)$. This is a reasonable assumption, since muon damped sources are typically connected with large magnetic fields and are expected at ultra-high energies, while neutrinos from neutron decay are typically produced at lower energies due to the kinematics of the neutron decay \cite{Hummer:2010ai}. For most astrophysical sources with not too strong magnetic fields there is therefore an intermediate energy range in which the pion beam regime is expected. Under this hypothesis, the flavor composition at the Earth will be obtained using the mixing probabilities given by Eq. (\ref{eq:average}). We let the parameters of the unitary matrix $\tilde{U}$ vary freely and we compare the expected flavor composition at the Earth with the likelihood contours obtained by the latest IceCube statistical analysis \cite{Abbasi:2020zmr} in Fig. (\ref{fig:flavtriang}). We also show the projected flavor sensitivity of IceCube-Gen2 \cite{Bustamante:2019sdb} under the assumption of pion beam sources: the sensitivity of IceCube-Gen2 to flavor composition has also been studied in \cite{Shoemaker:2015qul}.
The orange region is obtained by spanning the whole parameter space for the unitary matrix $\tilde{U}$: in particular, it includes the possibility that the gravitational basis coincides with the flavor eigenstates (green) or the mass eigenstates (black). In the latter case VEP averaged oscillations leads to the same result as conventional averaged oscillations, so that no constraints can be drawn from the observation of the flavor composition of astrophysical neutrinos: in fact, the black region cannot be excluded even by comparison with the projected IceCube-Gen2 sensitivity. On the other hand, in the case that gravitational and flavor bases coincide, VEP effects cause a suppression of the oscillations, as discussed above, so that the resulting flavor composition at the Earth is very far from the center of the triangle and therefore potentially subject to severe constraints by the future observations of IceCube-Gen2. The present information provided by IceCube is not generally able to significantly constrain the VEP parameters, since most of the orange region in Fig. \ref{fig:flavtriang} lies within the 95$\%$ confidence level of the experiment: however, it is noteworthy that a small number of choices, corresponding to the upward right corner of the orange region, are already outside the 95$\%$ contours, meaning that for such choices $\gamma_{ij} \phi>10^{-31}$ is already excluded at $2\sigma$ level. Interestingly, the flavor compositions which can be obtained under VEP span a rather vast area of the flavor triangle, exceeding the region obtained for example in Ref. \cite{Bustamante:2015waa}; it was in fact already noted by the authors that scenarios in which the oscillation mechanism is non standard are able to populate practically the entire triangle (see also \cite{Arguelles:2015dca}). In fact, the region in Fig. \ref{fig:flavtriang} has also been obtained as the general unitarity bound on the flavor compositions originating from pion beam sources in Ref. \cite{Ahlers:2018yom}: we emphasize that in our context each point of the region identifies a choice of the gravitational basis.

A particularly interesting observation can be made based on the recent observation by IceCube \cite{Abbasi:2020zmr} of astrophysical tau neutrinos. Since all the models of neutrino production \textit{via} hadronic interactions lead to electron and muon neutrinos only, the observation of tau neutrinos can be regarded as an unequivocal proof that neutrino oscillations are present during their propagation to the Earth. This statement has important consequences on the parameter space of VEP: in fact, if the gravitational basis coincides with the flavor basis, for $\gamma_{ij}\phi > 10^{-31}$ mixing between different flavors is inhibited by the VEP interaction, as noted above. Since no tau neutrinos are produced at the astrophysical source, this would lead to the conclusion that no tau neutrinos should be observed at the detector, a conclusion that has now been rejected with a statistical significance of nearly 96$\%$. As a consequence, in Fig. \ref{fig:flavtriang} the green region, corresponding to the case of gravitational basis coinciding with the flavor basis, is already excluded by the IceCube data at 95$\%$ confidence level. We remark that this is true independently of the assumption that the hadronic production is in the pion beam regime, and only comes from the the requirement that  oscillations are required to reproduce the observation of tau neutrinos.

\section{Conclusions}
The investigation of the foundations of general relativity has the potential to provide precious information for the construction of a fully consistent theory of gravitation. The study of VEP may therefore prove of crucial importance for future developments in this field. In this work, we analyzed the conclusions that can be drawn from high and ultra-high energy neutrinos on the possible strength of VEP. 

More specifically, in the energy range between 100 GeV and roughly 1 PeV, we analyzed the data samples recently released by IceCube, in particular IC79, IC86-11 and IC86-12/18. We studied the angular distribution of upgoing muon tracks using a $\chi^2$ methodology, taking into account the statistical and systematic uncertainties both on the IceCube data and on the atmospheric fluxes. We tested the hypothesis that VEP might exist which are diagonal in the mass basis, obtaining up-to-date constraints which are  stronger than the sensitivity estimated in Ref. \cite{Esmaili:2014ota}. We also discussed the dependence of our results on the choice of the gravitational basis in which VEP is diagonal, showing that for the extreme case in which this basis coincide with the flavor basis basically no constraints can be obtained from the analysis of atmospheric events. 

We analyzed the possible effects on astrophysical fluxes of VEP induced oscillations dominating over the conventional ones. This scenario can be achieved for values of $\gamma_{ij}\phi$ which are at present unconstrained by other experiments on neutrino oscillations: in particular, it requires that $\gamma_{ij}\phi>10^{-31}$. We show that in this regime the predictions for the flavor composition for simple models of astrophysical sources can be completely unrelated to the standard ones, depending on the choice of the gravitational basis. We find that for most choices the IceCube exclusion contours are not able to constrain to more than $1\sigma$ the VEP: however, we emphasize that for a limited number of choices, including the possibility that the gravitational basis coincides with the flavor basis, there is already a $2\sigma$ tension with the data. Our comparison shows that the projected flavor sensitivity of IceCube-Gen2 might be sufficient to exclude most of the possible choices of the gravitational basis in the scenario in which $\gamma_{ij}\phi>10^{-31}$.

We also comment on the interplay between the parameter space of VEP and the recent observation of astrophysical tau neutrinos by IceCube: the latter is in fact a smoking gun signature of the presence of oscillations during the neutrino propagation to the Earth. We find that this leads already with present data to the possibility of excluding at 95$\%$ confidence level the chance of a VEP diagonal in the flavor basis with $\gamma_{ij}\phi > 10^{-31}$.

Finally, we would like to highlight the complementarity between the two suggested approaches, using the IceCube data either in the atmospheric or in the astrophysical range. In fact, we find that the study of the atmospheric neutrinos will be nearly insensitive to the choice of a gravitational basis similar to the flavor ones. On the opposite, the study of astrophysical neutrinos will be mostly sensitive exactly for this choice, while being nearly insensitive to the choice of gravitational basis similar to the mass basis. We emphasize that, with the future data provided by IceCube-Gen2, a combined analysis of these two sources of data might be key in providing uniform and stronger constraints on the whole parameter space of VEP.

\subsection{Acknowledgments}
This work was supported by the research grant number 2017W4HA7S ``NAT-NET: Neutrino and Astroparticle Theory Network'' under the program PRIN 2017 funded by the Italian Ministero dell'Istruzione, dell'Universit\`a e della Ricerca (MIUR), and INFN Iniziativa Specifica TAsP.

\end{document}